# Enhancing AI Accessibility in Veterinary Medicine: Linking Classifiers and Electronic Health Records

**Chun Yin Kong[1], Picasso Vasquez[2], Makan Farhoodimoghadam[3], Chris Brandt[4], Titus C. Brown[5], Krystle L. Reagan[6], Allison Zwingenberger[7], Stefan M. Keller[1*]**

[1] Department of Pathology, Microbiology, Immunology, University of California Davis, USA
[2] Independent Researcher
[3] Department of Computer Science, University of California Davis, USA
[4] School of Veterinary Medicine Information Technology, University of California Davis, USA
[5] Department of Population Health & Reproduction, University of California Davis, USA
[6] Department of Veterinary Medicine and Epidemiology, University of California Davis, USA
[7] Department of Surgical & Radiological Sciences, University of California Davis, USA
[*] Corresponding author

**ORCID ID & email:**
Chun Yin Kong: 0009-0008-7731-9294, cykkong@ucdavis.edu
Picasso Vasquez: picassovasquez@gmail.com
Makan Farhoodimoghadam: mfarhoodi@ucdavis.edu
Chris Brandt: 0009-0005-0862-3629, cmbrandt@ucdavis.edu
Titus C. Brown: 0000-0001-6001-2677, ctbrown@ucdavis.edu
Krystle L. Reagan: 0000-0003-3426-6352, kreagan@ucdavis.edu
Allison Zwingenberger: 0000-0002-8982-2558, azwingen@ucdavis.edu
Stefan M. Keller: 0000-0002-5428-2985, smkeller@ucdavis.edu

## Abstract

**Background:** In the rapidly evolving landscape of veterinary healthcare, integrating machine learning (ML) clinical decision-making tools with electronic health records (EHRs) promises to improve diagnostic accuracy and patient care. However, the seamless integration of ML classifiers into existing EHRs in veterinary medicine is frequently hindered by the rigidity of EHR systems or the limited availability of IT resources.

**Results:** To address this shortcoming, we present Anna, a freely-available software solution that provides ML classifier results for EHR laboratory data in real-time. Anna is a standalone platform developed in Python, designed to host ML classifiers, retrieve patient-specific data from an EHR system, generate classifier results and return these results to the EHR for display. Anna merges results from different diagnostic tests according to user-defined temporal criteria and determines whether the data are sufficient for a given classifier. Because Anna is a stand-alone platform, it does not require substantial modifications to the existing EHR, allowing for easy integration into existing computing infrastructure. To demonstrate Anna's versatility, we implemented three previously published ML classifiers to predict a diagnosis of hypoadrenocorticism, leptospirosis, or a portosystemic shunt in dogs.

**Conclusion:** Anna is an open-source tool designed to improve the accessibility of ML classifiers for the veterinary community. Its flexible architecture supports the integration of classifiers developed in various programming languages and with diverse environment requirements. Anna facilitates rapid prototyping, enabling researchers and developers to deploy ML classifiers quickly without modifications to the existing EHR system. Anna could drive broader adoption of ML in veterinary practices, ultimately enhancing diagnostic capabilities and patient outcomes.

## Main Body

**Background**

Artificial intelligence has become increasingly popular in aiding clinical decision-making in veterinary medicine [1]. Various classifiers have been developed to predict the diagnosis of specific diseases based on laboratory data, such as hyperadrenocorticism [2], hypoadrenocorticism [3], leptospirosis [4] and portosystemic shunt [5] in dogs. Despite the availability of these machine learning (ML) classifiers, implementing ML classifiers in clinical practice faces several hurdles. The ML classifiers without a web-based graphical user interface (GUI) often require specific software environments and dependencies (**Supplementary Table 1**). Veterinarians without advanced computational skills may find installing and configuring the necessary software technically challenging, creating a barrier to adoption [6]. ML classifiers with a GUI alleviate these concerns, providing quick access, more accessible human-computer interactions, and greater user exposure, particularly when web-based. However, GUIs often require users to manually input many clinical parameters, such as blood work data, which is effort-intensive and prone to human error. In addition, users must ensure that clinical data units of the input data align with the requirements of the ML classifier and may need to apply value transformations, like grouping of continuous variables, converting concentration units, or standardizing breed information. Another potential limitation of classifiers, whether they employ a GUI or not, is that they only generate results when they are manually triggered by the user. If a clinician doesn't recognize a particular disease as a differential diagnosis, the ML classifier will not be triggered to analyze the patient data for prediction. Finally, running multiple classifiers on patient data requires separate, independent analyses for each ML classifier. Consequently, there is a pressing need for a computing environment that automates data analysis with ML classifiers. The analysis should seamlessly integrate with existing electronic health record (EHR) workflows and operate in the background without disrupting the clinician's routine. Crucially, this environment must be user-friendly, requiring no computer skills or specialized knowledge from the clinician.

To overcome these shortfalls, we developed Anna, an analytics tool that facilitates the integration of multiple ML classifiers into a single platform. Anna can retrieve patient data from an EHR system and generates ML classifier predictions without the need for manual data entry. This provides clinicians with just-in-time access to multiple prediction results. We demonstrate Anna's versatility by integrating three previously published ML classifiers for prediction of hypoadrenocorticism, leptospirosis and portosystemic shunt diagnosis in dogs. Anna is designed with a modular architecture, enabling the integration of future classifiers regardless of the programming languages used.

**Implementation**

1. **General architecture and software requirements**

Anna is a stand-alone platform that can host multiple ML classifiers and interfaces with EHRs to provide classifier predictions for patient data in real time. Upon request from an EHR system, Anna fetches relevant laboratory test results from the EHRs, runs available classifiers, and returns classifier prediction results to the EHR system for display to the user.

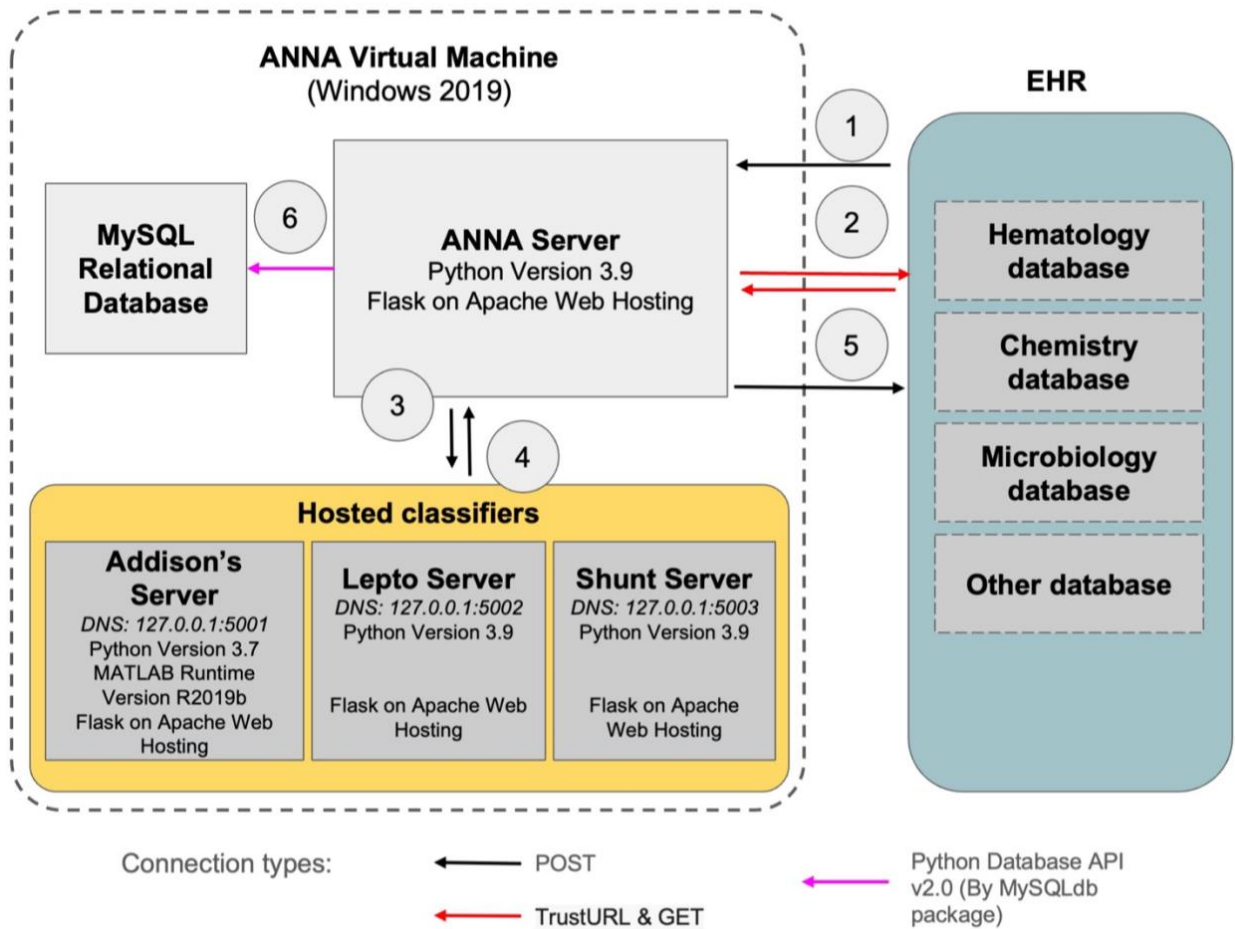

**Figure 1 ANNA architecture and workflow.** The prediction pipeline is triggered by a user visiting a specific test result in the EHR and comprises the following steps: 1) Classifier results request. The EHR requests a classifier result from the ANNA server by sending the patient ID and test date via REST API; 2) Data fetch. ANNA fetches relevant data from the EHR via TrustURL request and REST API; 3) Data pre-processing. ANNA merges results from different diagnostic tests according to user-defined temporal criteria and sends data to classifier servers; 4) Classifier servers generate prediction results and return the results to the ANNA server; 5) The ANNA server passes the classifier results to the EHR system for display; 6) ANNA stores classifier results with timestamps in a MySQL database.

Using end of support Python version is because Addison's, Lepto, and Shunt classifier are well-tested classifiers. We did not want to re-test them under a new version of Python.

Anna is a Python based server built with the Flask web framework and hosted using Apache HTTP Server services. It is a REST server that is not accessible via web browser and doesn't have a graphical user interface (GUI). Anna is contained in a virtual machine running on Windows Server 2019 and communicates with the EHR system via REST API, an application programming interface that is used to access a resource within another service or application with HTTP requests. The software packages used in Anna are open-source and Anna's source code and instructions are publicly available.
(https://github.com/ucdavis/ANNA-AnimalHealthAnalytics).

Machine learning classifiers are hosted in dedicated environments on separate Apache servers and communicate with the Anna core server through unique URL paths. This architecture allows preserving the software versions under which a ML classifier was developed while keeping the Anna core server up to date with respect to securities standards and bug fixes.

To demonstrate Anna's versatility, we integrated three previously published machine learning classifiers, each with distinct computational environment requirements: 1) "Addison's", a MATLAB-based ML classifier with a Python wrapper script, that employs adaptive boosting with decision trees for the prediction of hypoadrenocorticism in dogs [3], 2) "Lepto", a Python-based support vector machine (SVM) ML classifier for the prediction of leptospirosis in dogs [4], and 3) "Shunt", a Python-based extreme gradient boosting (XGBoost) ML classifier for the prediction of portosystemic shunt (PSS) in dogs [5]. All classifiers use at least two of the following data types: signalment, complete blood count, serum chemistry panel, urinalysis, leptospirosis testing request.

Several software libraries are essential for hosting these third-party classifiers on Anna. In general, Anna uses the Python data science package "pandas" [7] to handle data conversions from the EHR during data preprocessing. The Addison's classifier requires the MATLAB runtime [8], a standalone set of instructions that can execute compiled MATLAB applications or components, to establish connections between Python and MATLAB, and execute MATLAB scripts in Python environments. The advantage of using the MATLAB runtime is that it doesn't require an active MATLAB license, and it is readily available for download. The leptospirosis classifier utilizes Python data science packages such as 'scikit-learn' [9], and "numpy" [10] for further data preprocessing, calculations, and retrieving machine learning prediction results. The shunt classifier requires the "XGBoost" [11] Python package for loading the pre-trained model and generating predictions.

2. ML classifier prediction workflow

The classifier prediction pipeline consists of six stages: 1) classifier result request, 2) data fetch, 3) data preprocessing, 4) classifier prediction, 5) result and error reporting, and 6) results storage.

*2.1 Request for classifier result*
The generation of ML classifier results is initiated by a REST API request from the EHR system to Anna, containing the patient identification number and query date encapsulated in JSON format. Each classifier is associated with a unique URL. Triggers for initiating a request are flexible and may include a user activating a specific control or merely viewing a particular test result. In the current implementation, the trigger is viewing any laboratory test results that is required for a classifier where the patient species is "Canine" and the user is authorized to have access to the ML analysis.

*2.2 Data fetch*

To retain a lean architecture and minimize the chances of data loss, Anna only stores classifier results but no original patient data. Consequently, patient data is fetched from the EHR for each request. In the current implementation, each data fetch queries several EHR database tables (e.g. hematology, chemistry, microbiology, etc.) in parallel. The use of parallelism for data fetching significantly accelerates the execution of multiple queries as it is an I/O bound process. The EHR supports a REST-based process to request laboratory data and receive results in XML format when provided with a patient ID, date range, and laboratory section of interest. Access to this interface is restricted to specific IP addresses and requires an authorization code in order to respond to the request. Given that classifiers commonly depend on more than one type of laboratory test (e.g. CBC, serum chemistry panel and urinalysis) that may become available over several days, it is pertinent to capture test results that are reported in close temporal proximity. Anna fetches laboratory results within a user-defined time range from the query date. This time range is specific for each classifier and test type.

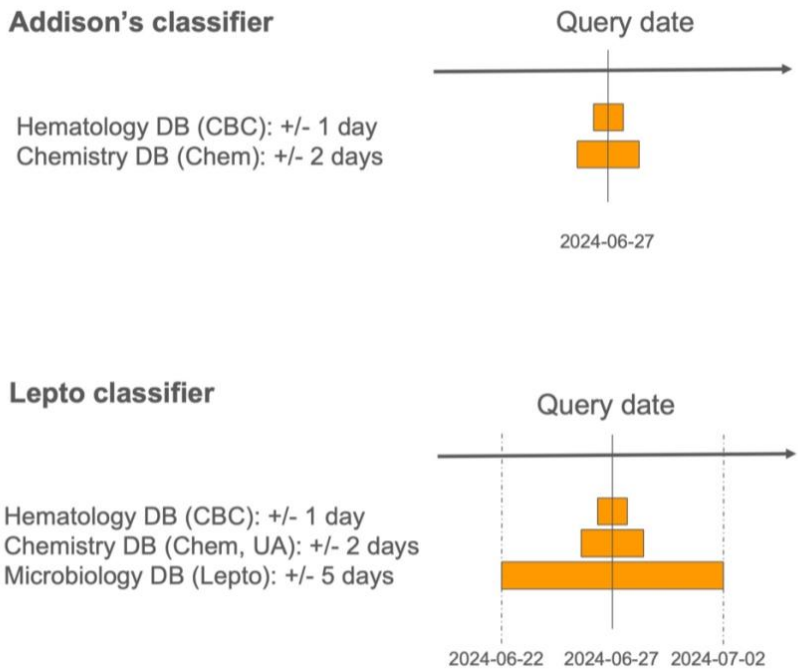

**Figure 2 Considerations for time range selection when retrieving data from the EHR.** To capture test results reported on days other than the query date, users can specify query time ranges for each database (DB) and classifier. The query date serves as the reference, and results are retrieved for the designated period before and after this date. For example: If the test date is "2024-06-27" (YYYY-MM-DD), the following date ranges will be fetched: Hematology DB: 2024-06-26 to 2024-06-28, Chemistry DB: 2024-06-25 to 2024-05-29, and Microbiology 2024-06-22 to 2024-07-02.

*2.3 Data Preprocessing*

The preprocessing step cleans data fetched from the EHRs system, merges individual tables into a master data frame and produces a classifier-specific data frame that is then passed to the ML classifier.

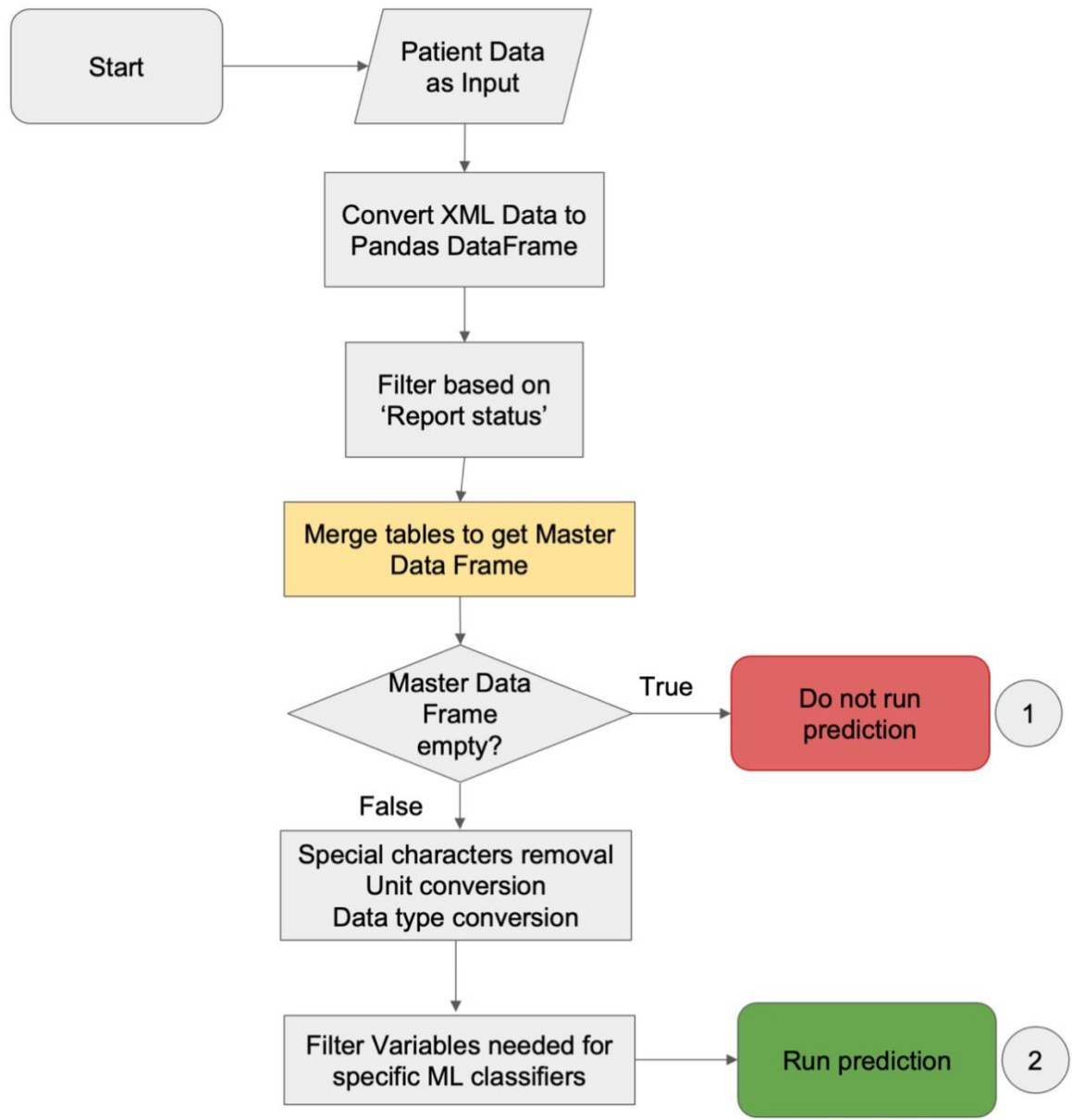

**Figure 3 Data preprocessing workflow.** The preprocessing merges tables fetched from several databases into a single data frame and performs filtering, data cleaning and conversion functions. If there is insufficient data, the classifier is not run and a result code of "-1" is returned to the EHR. If there is sufficient data, the master data frame is passed to the classifier server. Details of the 'table merging' step (highlighted in yellow) are described in Figure 4.

First, Anna filters records based on the report status. Typically, a finalized test result is preferred to ensure the accuracy and reliability of results that are passed to a ML classifier. However, in some instances, the test request, rather than the result, is of significance. For instance, the Lepto classifier is applied only when a diagnostic test for leptospirosis, such as Leptospira spp. PCR or microscopic agglutination test has been requested, as it was validated on suspected leptospirosis cases. Consequently, records containing either a finalized Lepto result or simply a test request are retained. Next, Anna merges the data, that were fetched and stored as separate tables, into a single master data frame. The merging logic is explained in **Figure 4**.

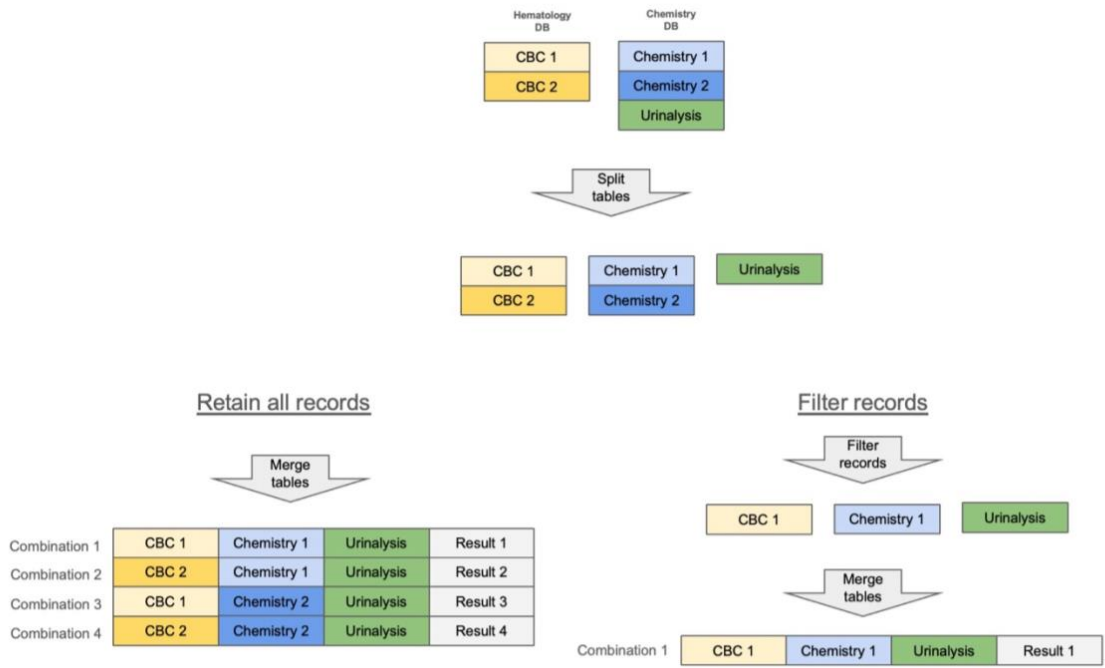

**Figure 4 Merging of test results.** If the data retrieved from a single database contains multiple test types (e.g. chemistry panel and urinalysis fetched from the Chemistry DB), then each test type is separated into its own table. Tables are then merged in an 'all-vs-all' fashion, i.e. if more than one result is available for a given test, then all possible test combinations are generated. For example, for a dataset with two complete blood counts (CBCs) and two chemistry panels (Chemistry), four unique combinations are created, and classifier results are generated for each combination. Alternatively, records can be filtered before table merging to limit the number of combinations generated based on a user-defined rule (e.g. only retain the first test per test type).

Subsequently, any special characters and string text, such as those found in comments, are eliminated from the numerical fields followed by the conversion of units (i.e. mg to µg) and data types (e.g. categorical to numerical data). Lastly, Anna selects and reorders the variables required by a given classifier and passes a classifier-specific data frame to the classifier module by serializing the data to JSON and sending it with POST requests.

## 2.4 ML classifier prediction and results storage

Classifier results are generated using a dedicated server for each classifier (**Figure 1**). This setup allows for a distinct computing environment for each classifier while ensuring security by restricting incoming requests solely to the Anna core server. For every line of the data frame, i.e. each unique set of test combinations, the classifier server generates the following information: 1) the prediction result (0/1 for binary classifiers, multi-level labels for multi-class classifier), 2) the IDs of tests that were used to generate the result, and 3) a timestamp reflecting when the ML classifier was first run on this particular data. This information is returned to the EHR database via JSON formatted string and stored in a MySQL database for legal purposes. Since classifier results might influence clinical decision making, they are considered part of the medical record and need to be archived. In the current implementation, changes to the EHR system were not possible, which required storage within Anna.

## 2.5 Error Reporting

Anna reports an error in two instances: 1) Insufficient patient data to run a given classifier or 2) an unsuccessful classifier prediction.

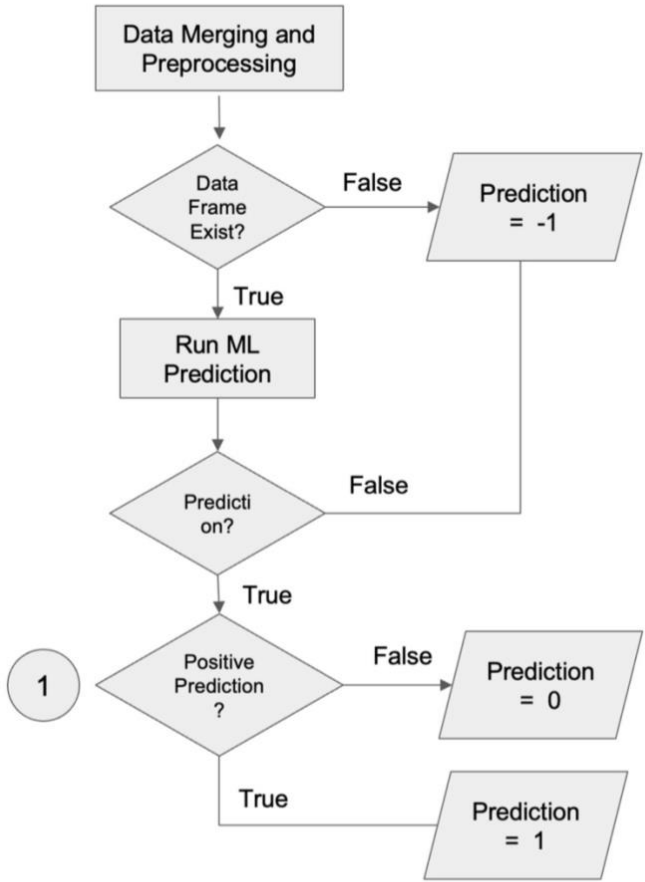

**Figure 5 Rules of ANNA returns error code to EHR.** 1) Decision outputs are based on binary ML classifiers. Prediction = 0 represents a negative prediction and prediction = 1 represents a positive prediction.

In both instances, Anna generates a prediction result of "-1" in the prediction field of the JSON string sent to the EHR system. In addition, Anna will store the error code and supporting information such as the start and end of a session, including intermediate steps with timestamps in log files locally. To identify each session, a random 6-digit ID is assigned so that Anna developers can quickly locate problems in the log files.

## 2.6 Results visualization

When the EHR receives the JSON response for the request, it presents the information to the clinical user with links to the specific tests that were used as the basis for its result. If multiple permutations of the results are returned due to repeat tests being performed within the test window, each possible option will be shown. Links to review detailed information on the specific ML classifier are also provided to help the user understand the purpose and mechanics of that specific classifier.

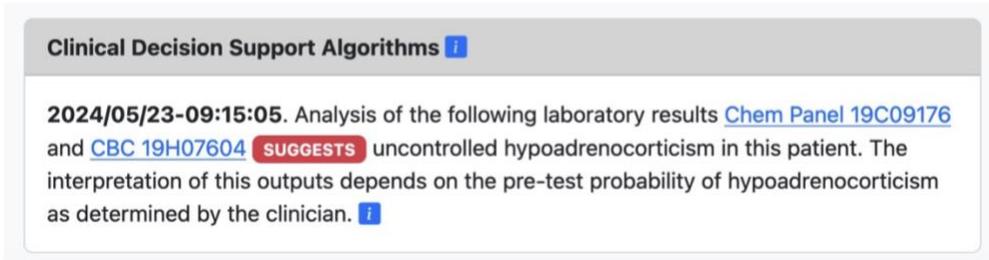

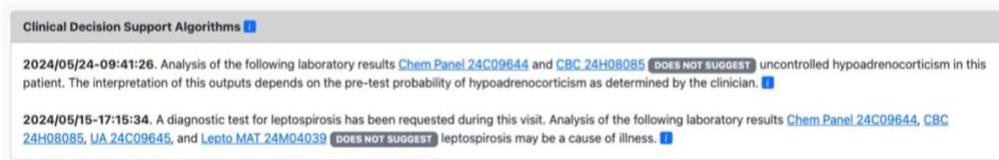

**Figure 6 Example of ML classifiers visualization in EHR.** A) Example of a positive classifier result as shown in the EHR. The timestamp reflects the date and time when the ML classifier was first run. The test IDs on which the result is based on are specified in the following sentence. B) Example of negative results for two different ML classifiers.

If the only results received by the EHR are "-1", it will display a message to the user "There are no eligible machine learning classifiers available for this patient."

3. **Anna Security Measures**

To protect the Anna core server from unauthorized access, incoming network traffic is controlled by two methods. First, Anna is accessible only within a specific network domain by routing through a virtual private network (VPN). Second, Anna authenticates the IP address of every incoming connection by cross-referencing the HTTP headers forwarded by the VPN service against an approved list of IP addresses. If an incoming IP does not match this list, the connection is terminated with a 403 Forbidden status, indicating the request is understood but not authorized.

Also, Anna developers can remotely control when EHR can communicate to each ML classifier's computation server through the Anna core server. Anna will refuse to respond to EHR requests with a 503 Service Unavailable status when the URL path is manually set to disabled.

To ensure secure communications between the Anna core server and the ML classifier servers, the DNS for each classifier server is set to 127.0.0.1, known as "localhost" or "loopback address." This keeps connections confined within the virtual machine. The Anna core server then makes local connections to each classifier server when receiving requests from the EHR system. Each classifier server is assigned a unique port number for distinction.

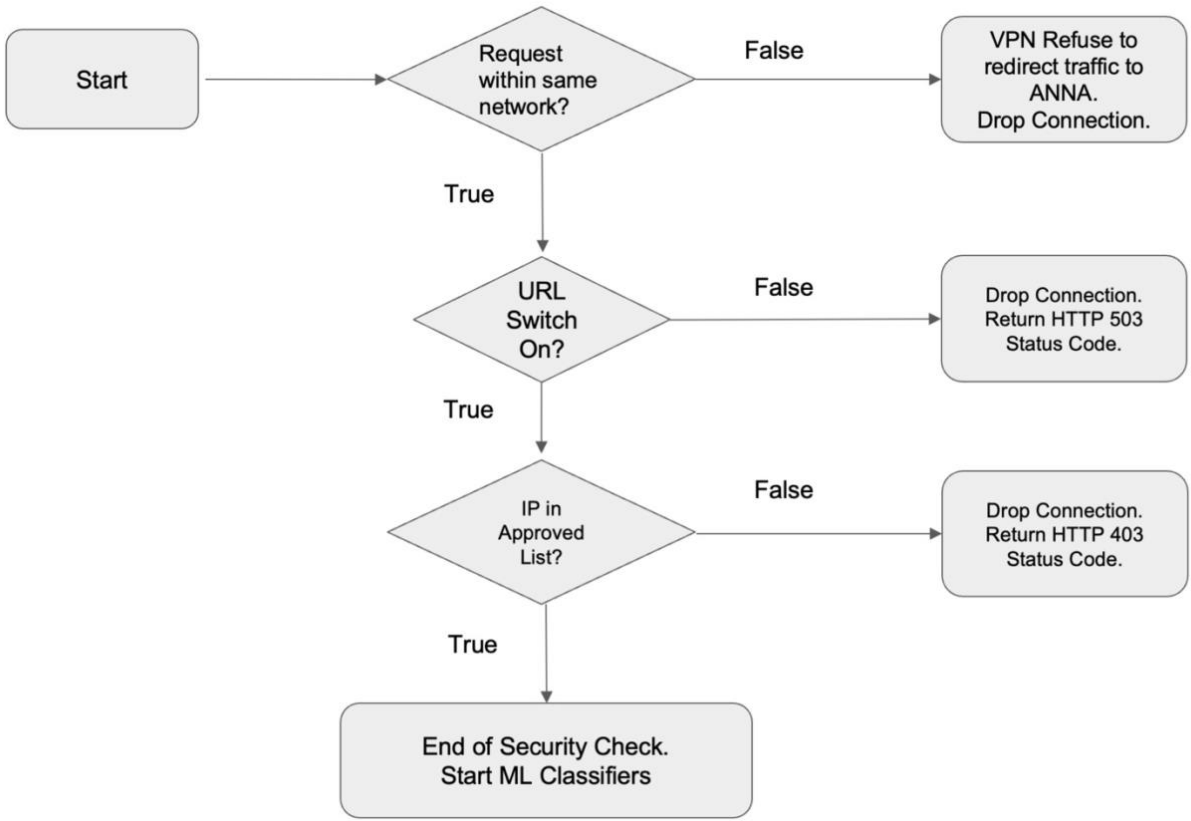

**Figure 7 Decision logic of the measures of protecting ANNA from unintended access.**

**Results/Discussion**

Herein we describe Anna, an animal health analytics platform, that provides the computing infrastructure to integrate ML classifiers with EHR systems for real-time analysis of veterinary laboratory data.

Integrating machine learning classifiers into existing EHR systems can be challenging due to resource limitations, lack of support, the rigidity of EHR systems or proprietary barriers. Since the UC Davis Veterinary Medical Teaching Hospital's EHR was developed in-house, modifications were generally feasible, but they needed to be minor in scope and require minimal IT resources. To meet these constraints, we shifted computations to a stand-alone platform and utilized existing reporting functionalities within the EHR. EHR modifications were limited to minimal changes to the export formats for lab results and the addition of the Anna server IP as a trusted system. The classifier results are currently displayed on the mobile version of the EHR for validation and will soon be integrated into the core EHR system. While Anna is designed for easy integration with EHR systems, it may face challenges with proprietary platforms that enforce strict limitations. Many commercial EHRs use closed architectures, restricting modifications or the addition of external tools like ANNA without significant involvement from the original developers.

The use of open-source software as the Anna backbone reduces cost and enhances accessibility for the broader veterinary community [12][13]. The rich support for multiple programming languages, third-party modules and plugins, and ability to reproduce in deployment helps Apache stand out as the ideal framework for hosting the backend server. Python's compatibility with different programming languages enables the integration of new ML classifiers, whether written in Python or not, without requiring significant code changes [14][15]. Additionally, its modular design allows for easy maintenance and future expansions with minimal disruption, as each ML classifier runs on its own dedicated server. Since Anna and the EHR are two separate systems and architectures, errors occurring in Anna will not affect the EHR, and vice versa. This ensures the stability of the EHR system as it is, in contrast to Anna, essential to the users in veterinary practices and clinics.

Hosting ML classifiers on dedicated servers improves performance and ensures the forward compatibility of classifiers. Since Anna is designed to host a diverse array of machine learning classifiers, we needed to create suitable environments to accommodate various programming languages and platforms. When integrating the hypoadrenocorticism classifier into Anna, we initially encountered prolonged prediction times compared to running it directly in MATLAB. This delay was primarily caused by Python initialization in the MATLAB runtime for every new request and could be resolved by a dedicated server running the ML classifier continuously. Implementing dedicated servers for ML classifiers also enables classifiers being run in legacy package versions with the dependencies used initially for compilation. This ensures the persistence of ML classifier results while keeping the Anna core server patched and updated. Independent servers can be easily set up in the existing environment without the need for additional hardware or software.

Anna simplifies obtaining predictions from multiple ML classifiers, overcoming the inefficiencies of current methods. Instead of visiting multiple sites for predictions or trying to set up classifiers locally, Anna fetches patient data directly from the EHR, runs the predictions on its backend server, and returns the results to the EHR for display. Users no longer need to manually input parameters or navigate different sites for various disease predictions, overcoming a major barrier for use in a clinical environment.

The visualization of prediction results is done by the EHR system and requires minimal modifications. After receiving the ML predictions in the form of a standardized JSON string, the EHR will display the results at the bottom of the test results that are currently viewed, which provides users with quick access to classifiers results without having to navigate to another site. At the same time, links to the specific tests that were used to make the prediction are provided.

Although veterinary medicine lacks data security laws like the Health Insurance Portability and Accountability Act (HIPAA) in human healthcare, securing patient data was a key focus in developing Anna. Healthcare data are sensitive since they are confidential [16] and usually come with personal identifiers that can be easily traced. Maintaining data privacy and preventing data leaks are two top concerns for adopting artificial intelligence into healthcare [17]. To address these concerns, Anna does not store patient data and relies on the EHR to control data access, keeping sensitive information secure. The only exception to this rule is the storage of the patient ID with test identifiers and timestamps in the open-source database management system MySQL. This became necessary because the classifier result is considered part of the medical record, but the EHR system of the Veterinary Medical Teaching Hospital at UC Davis does not yet provide a way to store this information. Yoon et. al introduced the concept of "accountability gap," where using

ML in clinical decisions can reduce human accountability, raising the question of responsibility for incorrect diagnoses [18]. As the adoption of AI into clinical practice becomes more commonplace, we anticipate that this information will be stored in the EHR rendering this functionality of Anna redundant.

Setting up Anna as a backend computational server offers three major benefits during development and deployment. First, during development, separating Anna from the EHR reduces the risk of conflicts with existing EHR code when adding and testing new features. Since EHR only handles presentation of the classifier result, fewer code changes are needed, and the risk of system crashes is lower compared to directly integrating ML classifiers into the EHR. Second, as a standalone server, Anna can be accessed by other EHR systems or authorized platforms without major code changes, thanks to its ability to control incoming traffic. Third, having separate teams working in parallel makes the integration progress smoother. This separation also allows for easier troubleshooting and maintenance, as issues related to Anna's operations can be resolved independently of the EHR system, minimizing disruptions to clinical workflows.

Future improvements of Anna should focus on including additional classifiers, facilitating the analysis of non-numerical data [19][20], providing interactive data visualizations, and making Anna accessible to the veterinary community. We have already successfully integrated classifiers written in both Python and MATLAB, demonstrating Anna's ability to work with various languages and configurations. Thanks to Anna's modular design and Python's flexibility to interact with classifiers written in different programming languages and utilizing various data types, Anna can be easily expanded to include additional ML classifiers in the future. We plan to expand Anna to display more information, such as time-series data and ML predictions over time and transform Anna with interactive visualization [21]. Lastly, to increase Anna's accessibility and reduce reliance on proprietary EHR systems, we want to enable users to upload data directly to Anna for analysis. This would bypass the need for EHR integration, making advanced machine learning accessible to the wider veterinary community, free of charge. This would not only lower the barrier to entry but also empower a more diverse group of practitioners to harness the potential of AI in their clinical work.

**Conclusions**
With the rapid expansion of artificial intelligence in veterinary medicine, the need for centralized, secure, and fast ML servers integrated with EHR systems is becoming increasingly urgent. Such integration reduces data retrieval times, minimizes human errors, and streamlines access to various ML classifiers. This paper discusses the development of Anna, a stand-alone centralized backend machine learning platform that facilitates real-time data analysis with ML classifiers, aiding in clinical decision-making. Future iterations of Anna will focus on integrating more third-party classifiers, expanding its capabilities to handle time-series patient data, enabling interactive visualizations, generating predictions from multi-visit data, and becoming publicly accessible for users to upload their own data for machine learning predictions.

**Availability and requirements**
    Project name: Anna/UC Davis Animal Health Analytics Platform
    Project home page: https://vmacs-analytics.vetmed.ucdavis.edu

Operating system(s): Platform independent
Programming language: Python
Other requirements: Apache, MATLAB, MySQL
License: GNU Affero General Public License
Any restrictions to use by non-academics: As defined by GNU Affero General Public License

**List of abbreviations**
- Anna - Animal Health Analytics Platform
- JSON - JavaScript Object Notation
- XML - Extensible Markup Language
- EHR - Electronic Health Record
- REST API - Representational State Transfer Application Programming Interface
- HTTP - Hypertext Transfer Protocol
- IP Address - Internet Protocol Address
- VPN - Virtual Private Network
- SQL - Structured Query Language
- I/O bound – Input/Output bound

**Citation**

## Supplementary Materials

| Classifiers | Species | Developers | GUI? | URL |
|---|---|---|---|---|
| Lepto Classifier | Dogs | Department of Veterinary Medicine and Epidemiology, University of California, Davis, CA 95616, USA | Yes | https://vmacs-analytics.vetmed.ucdavis.edu/ml_classifier_run/leptospirosis |
| Addison's Classifier | Dogs | Department of Veterinary Medicine and Epidemiology, University of California, Davis, CA 95616, USA | Yes | https://vmacs-analytics.vetmed.ucdavis.edu/ml_classifier_run/tommy_addisons |
| Shunt Classifier | Dogs | Department of Veterinary Medicine and Epidemiology, University of California, Davis, CA 95616, USA | Yes | https://vmacs-analytics.vetmed.ucdavis.edu/ml_classifier_run/shunt |
| Machine-learning based prediction of Cushing's syndrome | Dogs | Pathobiology and Population Sciences, The Royal Veterinary College, Hawkshead Lane, North Mymms, Hatfield, AL9 7TA Herts UK | Unknown | NA |

| Field identification of infectious and inflammatory disorders of the central nervous system in cattle | Cattle | Department of Veterinary Sciences, University of Turin, Italy | Yes | https://cnsprediction.streamlit.app/ |
|---|---|---|---|---|
| Myxomatous Mitral Valve Disease | Dogs | Department of Veterinary Internal Medicine, College of Veterinary Medicine, Seoul, Republic of Korea | No | NA |

**Supplementary Table 1 Table of ML Classifiers published to aid clinical-decision making in veterinary medicine.**